\documentclass[aps,pre,preprint,floatfix,twoside,tightenlines]{revtex4} 
\usepackage{amsmath,graphicx}

\DeclareMathOperator{\Ai}{Ai}
\DeclareMathOperator{\Bi}{Bi}
\DeclareMathOperator{\Tr}{Tr}
\DeclareMathOperator{\Det}{Det}

\DeclareMathOperator{\erf}{erf}
\DeclareMathOperator{\erfi}{erfi}
\DeclareMathOperator{\arccosh}{arccosh}

\newcommand{\order}[1]{\mathcal{O}(#1)}
\newcommand{\floor}[1]{\lfloor#1\rfloor}
\newcommand{\deriv}[3][]{\frac{\partial^{#1}#2}{\partial#3^{#1}}}
\newcommand{\fderiv}[3][]{\frac{\mathrm{d}^{#1}#2}{\mathrm{d}#3^{#1}}}

\begin{document}

\title{Quantum free energy differences from non-equilibrium path
  integrals: II.~Convergence~properties for the harmonic oscillator}

\author{Ramses van Zon$^{\ddagger}$, Lisandro Hern\'{a}ndez de la
Pe\~{n}a$^{*\dagger}$, Gilles H. Peslherbe$^\dagger$, and Jeremy
Schofield$^\ddagger$}

\affiliation{$^{\ddagger}$Chemical Physics Theory Group, Department of
Chemistry, University of Toronto, 80 Saint George Street, Toronto,
Ontario M5S 3H6, Canada}

\affiliation{$^*$Department of Chemistry, University of Illinois at
Urbana-Champaign, Urbana, Illinois 61801}

\affiliation{$^{\dagger}$Center for Research in Molecular Modeling and
Department of Chemistry and Biochemistry, Concordia University,
Montreal, Canada}

\date{August 29, 2008}

\begin{abstract}
Non-equilibrium path integral methods for computing quantum free
energy differences are applied to a quantum particle trapped in a
harmonic well of uniformly changing strength with the purpose of
establishing the convergence properties of the work distribution and
free energy as the number of degrees of freedom $M$ in the regularized
path integrals goes to infinity.  The work distribution is found to
converge when $M$ tends to infinity regardless of the switching speed,
leading to finite results for the free energy difference when the
Jarzynski non-equilibrium work relation or the Crooks fluctuation
relation are used. The nature of the convergence depends on the
regularization method. For the Fourier method, the convergence of the
free energy difference and work distribution go as $1/M$, while both
quantities converge as $1/M^2$ when the bead regularization procedure
is used.  The implications of these results to more general systems
are discussed.
\end{abstract}

\maketitle

\section{Introduction} 

In the preceding paper\cite{preceding}, a non-equilibrium path
integral method for computing free energy differences based on
combining the Jarzynski equality\cite{Jarzynski1,Jarzynski2} and the
Crooks fluctuation relation\cite{Crooks1,Crooks2} with the path
integral formulation of the quantum mechanical partition sum
\cite{Feynman1,Feynman2,Berne86,BerneThirumalai86} was presented. The
path integral representation of the canonical partition function is
based on mapping a quantum system at finite temperature onto a
classical system with additional degrees of freedom. A non-equilibrium
process can be carried out on this isomorphic classical system along a
well-defined trajectory in fictitious time.  The Jarzynski and Crooks
relations are valid for such a process, but only under the assumption
that the work distribution converges as the parameter $M$ of the
regularization procedure applied to the infinite dimensional path
integral goes to infinity.

A regularization procedure is needed because particles are represented
as objects with an infinite number of degrees of freedom in the path
integral formulation.  As a result, non-equilibrium dynamical
processes in this representation can lead to divergences in
non-physical quantities, such as the average total Hamiltonian of the
particle or the work performed on the system in the fictitious
process, which is a central quantity in the Jarzynski and Crooks
relations. A regularization procedure restricts the number of degrees
of freedom to a finite number $M$ and results in finite but
$M$-dependent estimators for quantities of interest.  In the
regularized path integral representation, one finds that the
expression for the work takes the form of the difference between two
quantities which diverge as $M\to\infty$.  Having an estimator for a
physical quantities that take the form of the difference between two
diverging quantities is not unusual in the context of path
integrals\cite{Berne86}, but this does make it important to establish
the convergence properties of all relevant estimators.
Furthermore, even when it can be demonstrated that the
regularization procedure leads to convergent results, the viability of
the non-equilibrium path integral method as a means of computing
quantum free energy differences is strongly dependent on the rate of
convergence of the regularized path integral to the exact quantum
result.  Neither the convergence nor the rate of convergence was
addressed in detail in Ref.~\onlinecite{preceding}, although strong
numerical evidence of convergence was presented for a quantum particle
in a quartic potential.

In this paper, the rate of convergence of different regularization
procedures is examined in detail for the special case of a quantum
harmonic oscillator.  Harmonic systems have the advantage of being
often amenable to analytical treatment, allowing for closed form and
exact solutions.  Here, the convergence of the regularization
procedure is studied in three regimes of the non-equilibrium process,
i.e., the quasi-static, the finite time and instantaneous switching.
The work distribution is computed for each of these regimes and its
convergence, as well as the convergence of free energy difference are
analyzed as the number of degrees of freedom goes to infinity.  It
will be shown that the free energy difference and the work
distribution converge for both the Fourier and the bead regularization
procedures\cite{preceding}, with the latter converging more quickly.

The paper is organized as follows: Section~\ref{system} presents a
brief overview of the method as it applies to the harmonic oscillator.
Section~\ref{explicit} contains the analysis of the convergence of the
free energy under different regularization schemes. The work
distributions will be determined using a generating function technique
explained in Sec.~\ref{generatingfunction}.  In
Sections~\ref{quasistatic}, \ref{finiteswitching} and
\ref{instantaneousswitching} the quasi-static, finite time and
instantaneous switching processes, respectively, are studied.  A comparison
between the non-equilibrium work distribution generated by the
fictitious dynamics and that generated with real time quantum dynamics is made in
Sec.~\ref{realquantum}.  The conclusions are given in
Sec.~\ref{conclusions}.

\section{Method and model system}
\label{system}

We consider a one-dimensional quantum system with a Hamiltonian
operator $\hat{H}(\lambda)=\hat{T}+\hat{V}$, with $\hat T=\hat
p^2/(2m)$ and $\hat V=V(\hat x,\lambda)=\frac12m\lambda\hat x^2$.
Here the potential energy $V$ depends on a control parameter
$\lambda$, which is equal to the square of the frequency $\omega$. The
canonical partition function of this system at an inverse temperature
$\beta$ is defined by
\begin{equation}
   Z(\lambda) = e^{-\beta F(\lambda)} = \Tr e^{-\beta \hat{H}(\lambda)}
\label{Zdef}
\end{equation}
and can be written as\cite{Feynman1,Feynman2}
\begin{equation}
   Z(\lambda) = \int\! \mathcal{D}x\:
                e^{-\frac{1}{\hbar}S[x,\lambda]},
\label{partitionFunction}
\end{equation}
where the integral is over closed paths $x(s)$ [i.e.,
$x(\beta\hbar)=x(0)$] and the Euclidean action $S$ is a functional of
$x$ given by
\begin{eqnarray}
   S[x,\lambda] = \int_0^{\beta\hbar}\! \mathrm ds  \left[ \frac12m
                  \Big(\fderiv{x}{s}\Big)^2 + \frac12m\lambda x^2 \right].
\label{euclideanAction}
\end{eqnarray}
Here and below the $s$ dependence of $x$ in integrals over $s$ will
always be implied.  For this one-dimensional harmonic system, the
quantum free energy is known to be exactly
\begin{equation}
  F(\lambda) = \beta^{-1}\log[2\sinh(\beta\hbar\omega/2)].
\end{equation}

The non-equilibrium path integral approach of computing free energy
differences uses a non-equilibrium process defined by a fictitious
dynamics in which $\lambda$ is changed from $\lambda_A=\omega_A^2$ to
$\lambda_B=\omega_B^2$ over a time $\tau$, while starting at canonical
equilibrium corresponding to $\lambda=\lambda_A$. This fictitious
dynamics is derived by introducing a new field $p(s)$ which is also
periodic in imaginary time, satisfying $p(s)=p(s+\beta\hbar)$, leading
to an expression that has the form of a classical partition function:
\begin{equation}
  Z(\lambda) =  C \int\! \mathcal{D}x\mathcal{D}p\:e^{-\beta
  H[x,p,\lambda]} ,
\label{Zpx2}
\end{equation}
where the fictitious Hamiltonian is given by
\begin{equation}
 H[x,p,\lambda] =
     \int_0^1\! \mathrm du  \Bigg[
      \frac{p^2}{2m} 
      +\frac12\kappa\Big(\fderiv{x}{u}\Big)^2
      +\frac12m\lambda x^2 \Bigg].
\label{scaledH}
\end{equation}
Here $u=s/(\beta\hbar)$ is a scaled imaginary time variable and the
string tension is
\begin{equation*}
\kappa= \frac{m}{\beta^2\hbar^2}.
\end{equation*}

We are interested in a Hamiltonian process, with equations of motion
\begin{subequations}
\begin{align}
  \deriv{x}{t}
    &= \frac{\delta H[x,p,\lambda]}{\delta p(u)}
       = \frac{p}{m}
       \label{dxdt}
       \\
         \deriv{p}{t}
	   &= -\frac{\delta H[x,p,\lambda]}{\delta x(u)}
      = \kappa\deriv[2]{x}{u}-m\lambda x,
\label{dpdt}
\end{align}
\end{subequations}
in which $\lambda$ is time-dependent, and satisfies the boundary
conditions $\lambda(0) = \lambda_A$ and $\lambda(\tau) =
\lambda_B$. Then, defining the fictitious work as
\begin{equation}
  W = H[x(\tau),p(\tau),\lambda_B]-H[x,p,\lambda_A],
\label{Wdef}
\end{equation}
the following identities were shown to hold\cite{preceding}
\begin{align}
  \big<e^{-\beta W}\big>_{\lambda_A}
= e^{-\beta\Delta F},
\label{JE}
\end{align}
and 
\begin{align}
  P_f(W) = e^{\beta W}e^{-\beta\Delta F} P_r(-W),
  \label{CFR}
\end{align}
where $\langle.\rangle_{\lambda_A}$ denotes an average over
non-equilibrium trajectories whose initial conditions are drawn from a
canonical distribution with $\lambda=\lambda_A$, $\Delta
F=F(\lambda_B)-F(\lambda_A)$, $P_f(W)$ is the probability density to
do an amount of work $W$ during the process that takes $\lambda$ from
$\lambda_A$ to $\lambda_B$ (the \emph{forward} process), and $P_r(-W)$
is the probability density to do work $-W$ during a similar process
that starts at $\lambda_B$ and ends at $\lambda_A$ (the \emph{reverse}
process). Equations \eqref{JE} and \eqref{CFR} are the Jarzynski's
non-equilibrium work relation and Crooks fluctuation relation,
respectively.

The distribution of the work done as a result of the change in
$\omega$ will be studied for this model.  This distribution can be
used either in the Jarzynski relation \eqref{JE} or in the Crooks
fluctuation relation \eqref{CFR} to determine the free energy
difference, both of which should yield $\Delta F = F_B - F_A =
\beta^{-1} \log \left[ \sinh(\beta\hbar\omega_B/2)
/\sinh(\beta\hbar\omega_A/2) \right]$.

The work in Eq.~\eqref{Wdef} is expressed as the difference of the fictitious
Hamiltonian at two times. The average value of the fictitious
Hamiltonian diverges in canonical equilibrium due to the infinite
number of degrees of freedom in the path integral
\eqref{partitionFunction}, which might pose a problem for the very
definition of the work distributions. To investigate whether the work
distribution is well defined, it is necessary to limit the system to a
finite number of degrees of freedom.  In Ref.~\onlinecite{preceding},
two different regularization methods of reducing the path integral
representations to a finite number of dimensions were introduced.
While both schemes are similar in their Fourier representations, in
one case the degrees of freedom correspond to low-frequency modes of
the continuous closed string, whereas in the other case they represent
a discretized lattice version of the string. In the first case the
regularization is based on statistical arguments motivated by the form
of the resulting Hamiltonian, and in the second case the
regularization is introduced in the lattice representation by means of
the Trotter formula.

For finite $M$, the value of $\Delta F$ found using the work
distribution differs from the exact quantum result. In fact, for the
harmonic oscillator, one can express the free energy explicitly as an
expansion in inverse powers of $M$, and thus assess the convergence of
$\Delta F$ analytically. This will be studied first in
Sec.~\ref{explicit}, in which alternative regularizations aimed at
improving the convergence are also discussed. Note that the
convergence of the free energy only requires equilibrium
considerations. Then, in Secs.~\ref{quasistatic},
\ref{finiteswitching} and \ref{instantaneousswitching}, the
convergence of the non-equilibrium work distributions is
analyzed for the cases of an infinitely slow switching rate, a finite
switching rate and an instantaneous switching process, respectively.

\section{The free energy under different regularizations}
\label{explicit}

Analytical considerations of the harmonic system proceed most easily
in the Fourier representation.  As explained in
Ref.~\onlinecite{preceding}, the Fourier transformation takes on
slightly different forms in the Fourier and the bead regularization
methods.  In both cases, though, the Hamiltonian assumes the form
\begin{equation}
H=\sum_{|k|\leq k_c} H_k,
\label{H}
\end{equation}
where $k_c$ is a cut-off wave vector, and
\begin{equation}
     H_k = \frac{|\tilde p_k|^2}{2m} +\frac12m\Omega^2_k|\tilde x_k|^2
\label{Hkdef}
\end{equation}
with
\begin{equation}
  \Omega^2_k = \omega_k^2+\omega^2.
\label{Omegandef}
\end{equation}
The dispersion relation for the Fourier regularization is given by
\begin{equation}
  \omega_k = 2\pi k\sqrt{\frac{\kappa}{m}} = \frac{2\pi k}{\hbar\beta},
  \label{dispersion-Fourier}
\end{equation}
while that for the bead regularization is
\begin{equation}
  \omega_k = \frac{2M}{\hbar\beta}\sin\frac{\pi k}{M}.
  \label{dispersion-beads}
\end{equation}
Note that in the Fourier regularization, $k_c$ is a chosen cut-off,
whereas from the Fourier transform of the bead regularization, we have
$k_c=\lfloor (M-1)/2\rfloor$ or, for $M$ odd, $M=2k_c+1$.
Thus, the two regularization methods can be parameterized either by
$k_c$ or $M$.

Given the Hamiltonian in Eq.~\eqref{Hkdef}, the partition sum may be
written as
\begin{align}
  Z_M(\lambda) = C\prod_{|k|\leq k_c}\int \mathrm d\tilde{p}_k \mathrm d\tilde{x}_k
  \: e^{-\beta H_k}
\end{align}
where $C$ is independent of $\lambda$.  Each integral is Gaussian and
can be explicitly evaluated. Since each mode occurs twice in the product (as
$k$ and $-k$) except for $k=0$, one finds for $e^{-\beta\Delta F}$,
\begin{align}
  e^{-\beta\Delta F}&=\lim_{M\rightarrow\infty}\frac{Z_M(\lambda_B) }{Z_M(\lambda_A)}
  \nonumber \\
  &= \lim_{k_c\rightarrow\infty} \frac{\omega_A}{\omega_B} \prod_{k=1}^{k_c}
  \frac{\omega_k^2+\omega_A^2}{\omega_k^2+\omega_B^2}.
\label{product}
\end{align}
This product must be evaluated separately for the two different
dispersion relations of the Fourier and the bead regularization
schemes in Eqs.~\eqref{dispersion-Fourier} and \eqref{dispersion-beads},
respectively.

In the Fourier regularization with the dispersion relation given by
Eq.~\eqref{dispersion-Fourier}, Eq.~\eqref{product} can be evaluated in the limit
$k_c \rightarrow \infty$ by writing
\begin{equation}
 e^{-\beta\Delta F}
= \frac{\omega_A}{\omega_B}
\prod_{k=1}^{\infty}
\biggl[1+\Bigl(\frac{\beta\hbar\omega_A}{2\pi k}\Bigr)^2\biggr]
\prod_{k'=1}^{\infty}
\biggl[1+\Bigl(\frac{\beta\hbar\omega_B}{2\pi k'}\Bigr)^2\biggr]^{-1},
\label{freeEn}
\end{equation}
and using the identity\cite{sinhIdentity}
\begin{equation*}
   \frac{\sinh z}{z} 
   = \prod_{k=1}^{\infty} \left(1 + \frac{z^2}{k^2\pi^2}\right),
\end{equation*}
to finally obtain the exact quantum result
\begin{equation}
 e^{-\beta\Delta F}
   = \frac{ \sinh(\beta \hbar \omega_A/2)}
          {\sinh(\beta \hbar \omega_B/2)}.
\label{Jarztest}
\end{equation}
It is straightforward to show using Eq.~\eqref{product} that the limit
is approached as $k_c^{-1}=\order{M^{-1}}$.

For the bead regularization, on the other hand, one uses
Eq.~\eqref{dispersion-beads} and $M=2k_c+1$ to write Eq.~\eqref{product} as
\begin{align}
  e^{-\beta\Delta F}
   & = \lim_{M\rightarrow\infty}
      \left[\prod_{k=1}^{M}
      \frac{\sin^2\frac{\pi k}{M}+\bigl(\frac{\hbar\beta\omega_A}{2M}\bigr)^2}
           {\sin^2\frac{\pi k}{M}+\bigl(\frac{\hbar\beta\omega_B}{2M}\bigr)^2}
\right]^{1/2}
\nonumber\\
   & = \lim_{M\rightarrow\infty}
      \left[\prod_{k=1}^{M}
      \frac{2+\bigl(\frac{\hbar\beta\omega_A}{M}\bigr)^2-2\cos\frac{2\pi k}{M}}
           {2+\bigl(\frac{\hbar\beta\omega_B}{M}\bigr)^2-2\cos\frac{2\pi k}{M}}
\right]^{1/2}.
\end{align}  
One then uses a different identity, namely\cite{LarsenRavndal88}
\begin{equation}
  \frac12\prod_{k=1}^M \Big(z - 2\cos\frac{2\pi k}{M}\Big)
  = \cosh\Big(M\arccosh \frac{z}{2}\Big)-1,
\label{identity}
\end{equation}
to arrive at 
\begin{align}
  e^{-\beta\Delta F}=  
\lim_{M\rightarrow\infty}
      \left[
      \frac{\cosh\{M\arccosh[1+\frac12\bigl(\frac{\hbar\beta\omega_A}{M}\bigr)^2]\}-1}
           {\cosh\{M\arccosh[1+\frac12\bigl(\frac{\hbar\beta\omega_B}{M}\bigr)^2]\}-1}
\right]^{1/2}
, 
\end{align}
which reduces to Eq.~\eqref{Jarztest} as well when the limit $M\to\infty$ is evaluated, with
correction terms of order $M^{-2}$ (see
e.g.\ Ref.~\onlinecite{LarsenRavndal88}).

There are other regularization methods possible which lead to even
faster convergence of the free energy at the expense of a more
complicated regularized Hamiltonian.  Such higher-order schemes can be
derived systematically by exploiting the analogy between the bead
representation and the Hamiltonian-splitting method used to obtain
integrators in molecular dynamics.  In fact, the basis of the bead
regularization is the splitting form in Eq.~(30) of the preceding
paper\cite{preceding}, and that same form is also the basis of the
Verlet scheme to integrate the equations of motions for classical
Hamiltonian systems in molecular dynamics
simulations\cite{FrenkelSmit}. The advantage of using splitting
schemes to derive integrators for molecular dynamics is that the
approximate dynamics is still symplectic, causing such integration
schemes to be very stable. The analogous property to symplecticity in
the path integral context is the Hermitian nature of the Boltzmann
operator, which is preserved in splitting schemes.

In an attempt to reduce the error due to operator splitting, many
alternative splitting schemes for molecular dynamics simulations have
been derived (see Ref.~\onlinecite{Omelyanetal03} and references
therein). Some of these schemes raise the order of the splitting
approximation to $\order{\delta t^4}$ or higher. Such splitting
schemes can also be used for path integrals and result in a fictitious
Hamiltonian in which the beads are not all equivalent or which
contains explicit correction terms of order $\order{\hbar^2}$ in the
potential, and whose partition function converges to the real quantum
partition sum as $\order{M^{-4}}$ or higher.  Other splitting schemes
derived for molecular dynamics simulations are aimed at reducing the
error by minimizing the pre-factors in front of the leading correction
terms\cite{Omelyanetal02}. These methods, however, are based on the
assumption that different correction terms contribute independently
and equally to the error.  This approach has proved useful in
molecular dynamics simulations.  For instance, it has been
demonstrated that an optimized second order scheme called HOA2 can
often outperform the Verlet scheme in molecular dynamics simulations
of rigid water molecules\cite{VanZonetal08a}.

We will briefly consider the HOA2 operator-splitting scheme applied as
a means to regularize imaginary-time path integrals, since it is the
simplest splitting-method that may improve the rate of convergence of
the regularization. It is based on the following operator-splitting
scheme:
\begin{equation}
e^{-\beta\hat H/M} = 
e^{-\eta\beta\hat V/M}
e^{-\beta\hat T/(2M)}
e^{-(1-2\eta)\beta\hat V/(2M)}
e^{-\beta\hat T/(2M)}
e^{-\eta\beta\hat V/M}
+\order{M^{-3}} .
\label{hoa2}
\end{equation}
For $\eta=1/4$, this splitting scheme reduces to the path
integral analog of the Verlet scheme of molecular dynamics (applied
twice), which corresponds to the standard bead regularization
procedure.  However taking a different value of $\eta$, namely
$\eta=0.1931833275037836$, minimizes the
sum of the squares of the error terms of
$\order{M^{-3}}$\cite{Omelyanetal02}.  It is shown in Appendix
\ref{appA} that when applied to the dynamics of a classical harmonic
oscillator, this scheme conserves energy better than the Verlet
scheme. The application of the HOA2 splitting scheme to path
integrals is 
carried out by representing the Boltzmann operator by $M$ factors of
$\exp(-\beta\hat H/M)$, 
taking its trace to get the partition function,  using the splitting
scheme \eqref{hoa2} for each factor, inserting
completeness relations between each exponential, and performing the
momentum integration. This procedure leads to a regularized path integral with
Hamiltonian
\begin{equation}
  H_M =  \sum_{n=1}^{M} \bigg\{
      \frac{mM}{\hbar^2\beta^2}\left[(v_n-u_n)^2
      +(u_{n+1}-v_n)^2\right]
      + \frac{1}{2M}\left[w_1 U(u_n)+w_2U(v_n)\right] \bigg\}.
\label{pairH}
\end{equation}
Here, $w_1=4\eta$, $w_2=2-4\eta$ and the $u_n$ and $v_n$ are the
positions of the odd and the even beads, respectively. Odd and
even beads are no longer identical in nature when $\eta \neq 1/4$
because the completeness relation
inserted either in front of the $e^{-(1-2\eta)\beta\hat V/(2M)}$ or the
$e^{-\eta\beta\hat V/M}$ in Eq.~\eqref{hoa2} gives rise to different
potential strengths $w_2$ and $w_1$. The partition sum
$Z_M=\int\!\mathrm d^Mu\,\mathrm d^Mv\:\exp(-\beta H_M)$ is of 
Gaussian form, since $H_M$ can be written as $H_M=\Gamma^T\mathsf
V\Gamma$, with $\Gamma=(u_1,v_1,u_2,v_2,\ldots)$. The integral can be
evaluated in terms of the determinant of the matrix $\mathsf V$. For
the Verlet-like Hamiltonian, a Fourier transform could be used to
diagonalize the matrix (i.e., to decouple the modes), which
facilitates the determination of the eigenvalues and thus the
determinant of $\mathsf V$. Here, a Fourier transform only yields a partial
diagonalization, i.e., defining
\begin{align}
  \begin{pmatrix}
    \tilde u_k \\
    \tilde v_k 
  \end{pmatrix}
&= \frac{1}{\sqrt2M}\sum_{n=1}^M e^{-2\pi \mathrm ikn/M} 
  \begin{pmatrix} 
    u_n\\
   e^{-\pi \mathrm ik/M} v_n
  \end{pmatrix}
\label{pairFT}
\end{align}
(where the factor $\sqrt{2}$ and the shift in the phase in front of $v_n$ are
introduced for convenience),
one gets
\begin{equation}
  H_M = \sum_{k=1}^M \begin{pmatrix}\tilde u_k^*&\tilde v_k^*\end{pmatrix}
\cdot
\mathsf W_k
\cdot
\begin{pmatrix}\tilde u_k\\\tilde v_k\end{pmatrix}
\end{equation}
with
\begin{equation*}
  \mathsf W_k = \frac{4M^2}{\beta^2\hbar^2}
\begin{pmatrix}   2+w_1\left(\frac{\beta\hbar\omega}{2M}\right)^2
                & -2\cos\frac{\pi k}{M}
              \\ -2\cos\frac{\pi k}{M} & 2+w_2\left(\frac{\beta\hbar\omega}{2M}\right)^2
\end{pmatrix}
\label{Wk}. 
\end{equation*}
Since each term in the Hamiltonian is simply a $2\times2$ quadratic
form, the partition sum can be written as a product of two-dimensional
Gaussian integrals, each of which is proportional to $1/\sqrt{\det
\mathsf W_k}$. The finite-$M$ partition sum thus becomes
\begin{equation}
  Z_M =\prod_{k=1}^{M}\frac{1}{\left[ 2+4\big(\frac{\beta\hbar\omega}{2M}\big)^2
+w_1w_2\big(\frac{\beta\hbar\omega}{2M}\big)^4
-2\cos\frac{2\pi k}{M}
\right]^{1/2}}
\label{ZN5} .
\end{equation}
Using the identity in Eq.~\eqref{identity} gives the result
\begin{equation}
Z_M  =
\frac{1}{\sqrt{2\left[\cosh\big\{M\arccosh\big[ 1+\frac12\big(\frac{\beta\hbar\omega}{M}\big)^2
+\frac{w_1w_2}{32}\big(\frac{\beta\hbar\omega}{M}\big)^4\big]\big\}-1\right]}}.
\end{equation}
By expanding $Z_M$ in $M$, one can write this as
\[
Z_M =
\frac{1}{2\sinh(\beta \hbar \omega/2)} \bigg[ 1
+ \frac{z(\eta) (\beta\hbar\omega/2)^3}{\tanh(\beta\hbar\omega/2)
M^2} 
 + O(M^{-4})\bigg] ,
\]
where the prefactor for the first correction term is $z(\eta) = 1/6 -
\eta + 2 \eta^2 $.  Thus, different values of $\eta$ lead to different
convergence properties since $z(\eta)$ depends explicitly on $\eta$.
The leading order correction is minimized by choosing $z(\eta)=0$,
i.e., $\eta = 1/4$, which corresponds to the standard bead
regularization procedure.  Other choices for $\eta$ lead to larger
correction terms and hence slower convergence.

We therefore conclude that using a
different splitting scheme can be useful for path integrals if the
order of the approximation is changed, but that optimized splitting
methods need not yield any improvement, even if they have been shown
to be beneficial in the context of molecular dynamics simulations. We
will therefore work only with the Verlet-like Hamiltonian in the
remainder of this paper.

\section{Generating function of the work distribution}
\label{generatingfunction}

The calculation of the work distribution proceeds most easily by first
determining its generating function (which coincides with its Fourier
transform)
\begin{align}
  G(u) &= \int_{-\infty}^\infty\!\mathrm dW\: e^{\mathrm iuW} P(W)
\label{generating0}\\
       &=
\big<e^{\mathrm iu[H(\tau)-H(0)]}\big>_{\omega_A},
\label{generating}
\end{align}
where $H(t)= H(\tilde{\mathbf x}(t),\tilde{\mathbf p}(t),\lambda(t))$, which
in this case can be written as in Eq.~\eqref{H}.

\begin{figure}[t]
\centerline{\includegraphics[width=0.9\columnwidth]{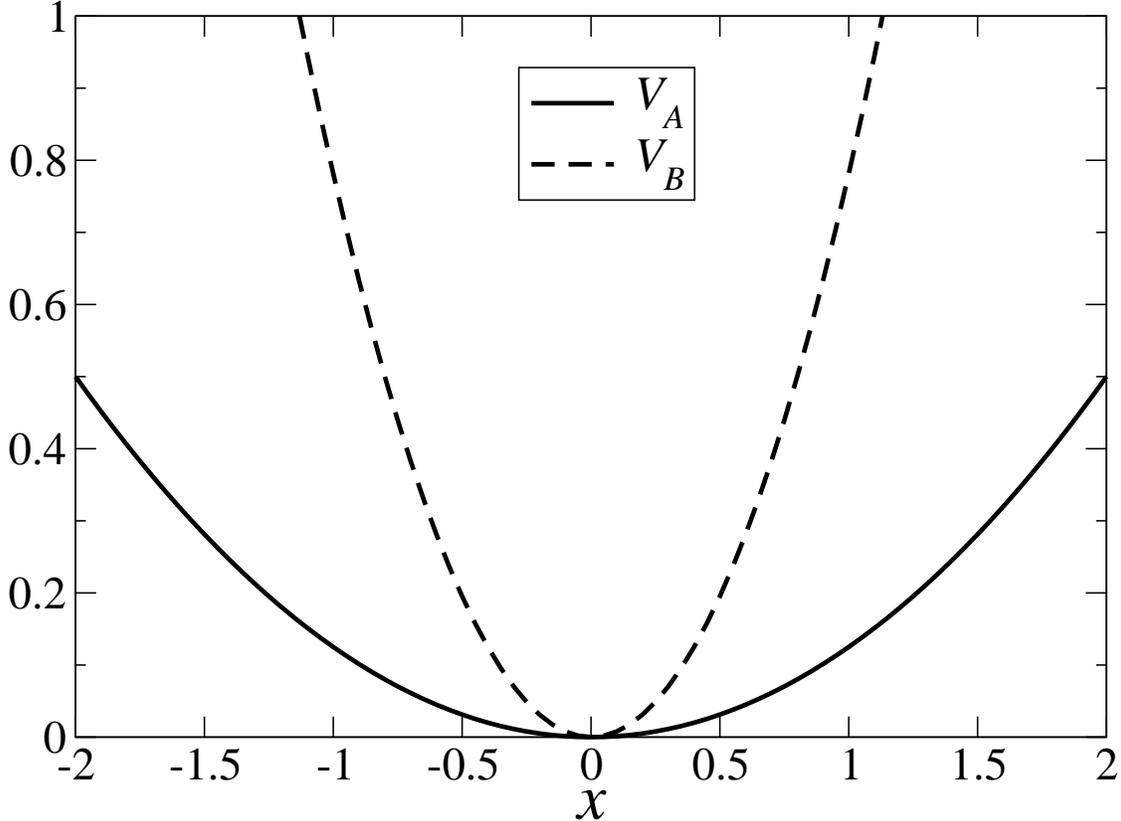}}
\caption{The two harmonic potentials $\frac12m\omega^2x^2$ for which
         the free energy difference and the work distribution in
         switching from one to the other is studied. In both cases,
         $m=1$, while for $V_A$, $\omega=\omega_A=1/2$ and for $V_B$,
         $\omega=\omega_B=5/4$. }
\label{fig:switch-oscillator}
\end{figure}

The equations of motion are given by
\begin{subequations}
\begin{align}
  \fderiv{\tilde x_k}{t}   =& \frac{\tilde p_k}{m}
\label{eom1}
\\
  \fderiv{\tilde p_k}{t}   =&  -m\Omega^2_k(t)\tilde x_k.
\label{eom2}
\end{align}
\end{subequations}

{}From the equations of motion, it is apparent that all Fourier modes
evolve independently with a time dependent frequency, and that each
mode contributes an independent term
\begin{equation}
W_k = H_k(\tau)-H_k(0)
\end{equation}
to the total work $W=\sum_{|k|\leq k_c} W_k$.  Furthermore, the modes
are also independent in the initial canonical distribution function
$\exp[-\beta H(\tilde{\mathbf x},\tilde{\mathbf p},
\omega_A^2)]$. Because the generating function of the sum of
independent term is the product of the generating functions of the
different terms, we get for $G(u)$
\begin{equation}
  G(u) = \prod_{|k|\leq k_c} G_k(u)
\label{Gu}
\end{equation}
where
\begin{align}
  G_k(u) &=
 \frac{\int \mathrm dx_k\, \mathrm dp_k\:
          e^{\mathrm iuW_k-\beta H_k(0)}}
{\int \mathrm dx_k\, \mathrm dp_k\:
          e^{-\beta H_k(0)}}.
\label{generatingn0}
\end{align}
Below, we will investigate the convergence of the work distribution functions
$P_f$ and $P_r$ as $M$ is taken to infinity.

\section{Quasi-static process}
\label{quasistatic}

\subsection{Work distribution}

Consider first a process in which $\omega$ is changed
infinitely slowly or quasi-statically from $\omega_A$ to $\omega_B$.
Since each mode $k$ is equivalent to a classical harmonic oscillator
with frequency $\Omega_k$, and $\Omega_k$ changes from
$\Omega_k(0)=\sqrt{\omega_k^2+\omega_A^2}$ to
$\Omega_k(\tau)=\sqrt{\omega_k^2+\omega_B^2}$, one can use that
$H_k(t)/\Omega_k(t)$ is an adiabatic invariant for harmonic
oscillators\cite{Goldstein}, to obtain for the work
\begin{equation}
W_k=[\Omega_k(\tau)/\Omega_k(0)-1]H_k(0).
\end{equation}
Eq.~\eqref{generatingn0} then gives for the generating functions of mode~$k$
\begin{align}
   G_k(u) &= 
   \frac{1}{1-\mathrm iu/\gamma_k} ,
\label{Gnu}
\end{align}
where
\begin{equation}
  \gamma_k 
      =\beta\left[
    \sqrt{\frac{\omega_k^2+\omega_B^2}{\omega_k^2+\omega_A^2}}-1
\right]^{-1}.
\label{andef}
\end{equation}
Note that all $\gamma_k$
have the same sign as $\omega_B-\omega_A$, cf.~Eq.~\eqref{andef}. 
The inverse Fourier transform of Eq.~\eqref{Gnu} is 
\begin{equation}
  P_k(W) = |\gamma_k| e^{-\gamma_kW}\Theta(\gamma_k W),
\end{equation}
where $\Theta$ is the Heaviside step function.  The
inverse Fourier transform of $G(u)$ in Eq.~\eqref{Gu} is a convolution of
these exponential functions, which yields
\begin{equation}
  P(W) = \Theta(\gamma_0W) \sum_{k=0}^{k_c} \Gamma_k(W) e^{-\gamma_kW},
\label{PW}
\end{equation}
where the form of $\Gamma_k (W)$ depends on whether mode $k$ is
degenerate or not.  For degenerate modes with $a_k=a_{-k}$ (i.e.\
$1\leq k \leq \floor{\frac{M-1}{2}}$), $\Gamma_k(W)$ is a linear
function of $W$,
\begin{subequations}
\begin{equation}
  \Gamma_k(W) = 
       \frac{|\gamma_k|}
        {\prod_{\stackrel{\mbox{\scriptsize$|q|\leq k_c$}}{|q|\neq k}} 
        \big(1-\frac{\gamma_k}{\gamma_q}\big)}
        \Bigg(
         \gamma_k W + \mathop{\sum_{|q|\leq k_c}}_{|q|\neq k}\frac{1}{1-\frac{\gamma_q}{\gamma_k}}
        \Bigg),
\label{Gamma-degenerate}
\end{equation}
while for non-degenerate modes ($k=0$, and $k=\frac M2$ if
$M$ is even), $\Gamma_K(W)$ is a constant given by
\begin{equation}
  \Gamma_k(W) = 
       \frac{|\gamma_k|}
        {\prod_{\stackrel{\mbox{\scriptsize$|q|\leq k_c$}}{q\neq k}} 
        \big(1-\frac{\gamma_k}{\gamma_q}\big)}.
\label{Gamma-non-degenerate}
\end{equation}
\end{subequations}

\begin{figure}[t]
\centerline{\includegraphics[width=0.9\columnwidth]{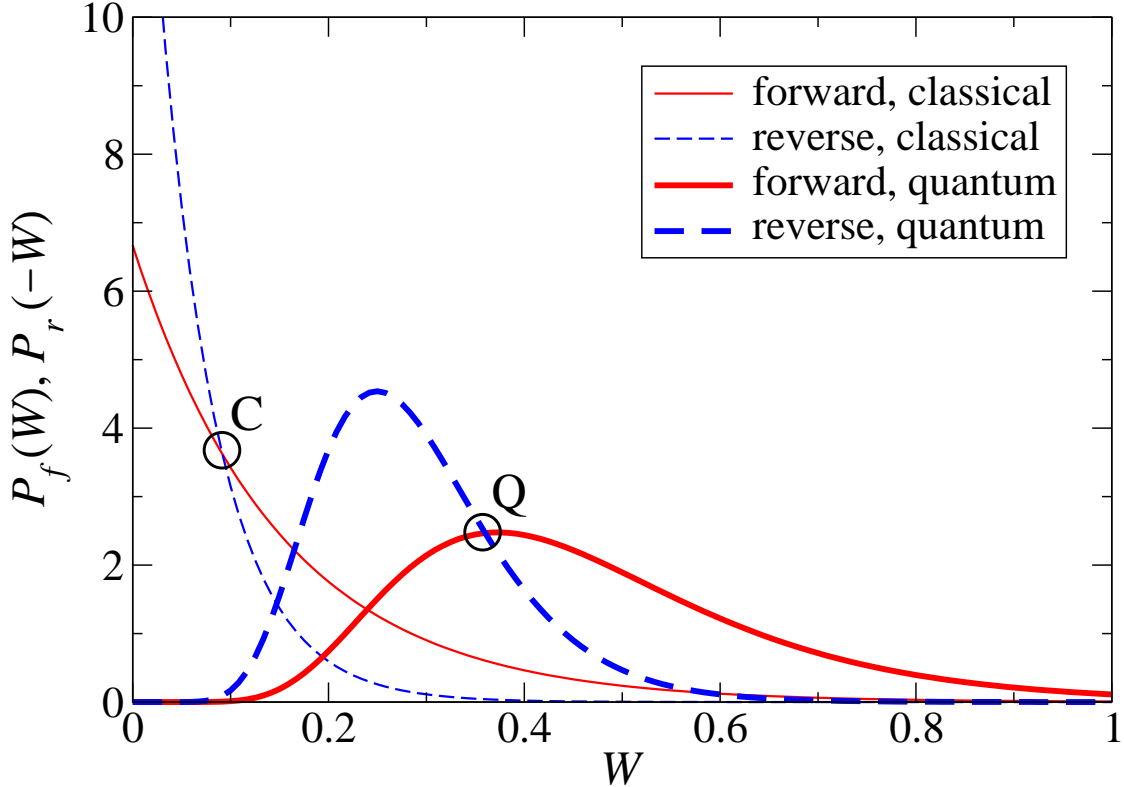}}
\caption{(Color online) Classical and quantum forward ($P_f$) and reverse ($P_r$)
         work distributions for a quasi-static switching between the
         potentials in Fig.~\ref{fig:switch-oscillator}, generated
         from Eq.~\eqref{PW} using the bead dispersion relation
         \eqref{dispersion-beads}, with $M=0$ (classical case) and
         $M=15$ (quantum case, converged up to $\approx5\%$) The
         circles labeled C and Q are the crossing points of the
         classical and the quantum distributions, respectively, whose
         $W$ values should coincide with the classical and quantum
         free energy differences, according to the Crooks fluctuation relation
         \eqref{CFR}. }
\label{fig:ho-work-distributions}
\end{figure}

As an example, let the
frequency of the harmonic oscillator be switched from $\omega_A=1/2$
to $\omega_B=5/4$, while setting
$m=\hbar=1$ and $\beta=10$ (see Fig.~\ref{fig:switch-oscillator}).  
In Fig.~\ref{fig:ho-work-distributions},
the forward and reverse work distributions are plotted, using Eq.~\eqref{PW}
with $M=0$ and $M=15$, where, for the reverse process, the values of
$\omega_A$ and $\omega_B$ are interchanged. The value of $M=15$ was
chosen because the work distribution has then already converged up to
about 5\%, while $M=0$ corresponds to the classical process. As
explained in the preceding paper\cite{preceding}, the work values at
crossing points of the forward and reverse distributions should be
equal to the free energy difference. In
Fig.~\ref{fig:ho-work-distributions}, the classical distributions are
seen to cross at $W_c\approx0.09$, which agrees with the prediction
$\Delta F_{\rm classical}=\beta^{-1} \log(\omega_B/\omega_A) =
0.0916\ldots$, while the quantum distributions cross at
$W_c\approx0.36$, which agrees with the prediction $\Delta F_{\rm
quantum} = \beta^{-1}\log ( \sinh(\hbar\beta\omega_B/2) /
\sinh(\hbar\beta\omega_A/2) )=0.375\dots$ within 4\%.  Better
agreement for the quantum case is obtained by using higher values of
$M$.

\subsection{Convergence of the work distribution}

Although the distribution functions in
Fig.~\ref{fig:ho-work-distributions} suggest a numerical convergence,
one can prove analytically that they converge by analyzing the
cumulants of the distributions, rather than the somewhat cumbersome
infinite products and sums in Eqs.~\eqref{Gamma-degenerate} and
\eqref{Gamma-non-degenerate}.  The cumulants $\kappa_j$ of the
distribution $P(W)$ follow from the generating function as
\begin{align}
\kappa_j=\left[\fderiv{}{(\mathrm iu)}\right]^j\log G(u)\bigg|_{u=0},
\end{align}
which means the generating function can be expressed in terms of
cumulants as
\begin{align}
   \log G(u)  
   = \sum_{j=1}^{\infty}\frac{\kappa_j}{j!}(\mathrm iu)^j.
\end{align}
The cumulants are therefore also formally related to the free energy difference
by the Jarzynski relation \eqref{JE}, i.e.,
\begin{align}
 \Delta F &= -\frac1\beta\log\langle e^{-\beta W}\rangle
 = -\frac1\beta\log G(\mathrm i\beta)
= \sum_{j=1}^\infty\frac{\kappa_j}{j!} (-\beta)^{j-1}.
\label{JEcum}
\end{align}
Cumulants of independent variables are additive, so that
\begin{equation}
  \kappa_j=\sum_{|k|\leq k_c} \kappa^{(k)}_j
   = \sum_{|k|\leq k_c} \frac{(j-1)!}{\gamma_k^j}
   .
\label{cums}
\end{equation}
where the mode cumulants were determined  from Eq.~\eqref{Gnu}.
The convergence of the cumulants as $k_c\to\infty$ can be determined
now.  From Eq.~\eqref{andef}, one sees that
\begin{equation}
   \gamma_k 
      \stackrel{k\gg1}\longrightarrow 
      \frac{2\beta\omega_k^2}{\Delta\omega^2}
   ,
   \label{gamma-asymptotics}
\end{equation}
where
\begin{align}
  \Delta\omega^2 = \omega_B^2-\omega_A^2,
  \label{delta-omega-def}
\end{align}
and therefore
\begin{equation}
   \kappa^{(k)}_j 
       \sim (j-1)!
      \left(\frac{\Delta\omega^2}{2\beta\omega_k^2}\right)^j 
   .
   \label{kappakj}
\end{equation}
The effect of this asymptotic formula is different in the Fourier and
in the bead regularization. In the Fourier regularization,
$\omega_k\sim k$, so that $\kappa^{(k)}_j\sim1/k^{2j}$. Thus one sees
that not only do all cumulants in Eq.~\eqref{cums} converge as
$k_c\to\infty$, higher order cumulants converge faster than lower
orders, i.e.\ 
\begin{equation}
\kappa_j(k_c)-\kappa_j(\infty) =
\order{1/k_c^{2j-1}}=\order{1/M^{2j-1}}.
\label{asymptote-Fourier}
\end{equation}
This means that the shape
of the work distribution converges faster than the average. However,
this property turns out not to be robust. One can show that adding a
perturbative quartic term to the potential causes \emph{all} cumulants
to converge as $k_c^{-1}$ in the Fourier
regularization\cite{unpublished08}.

\begin{figure}[t]
\centerline{\includegraphics[width=0.9\columnwidth]{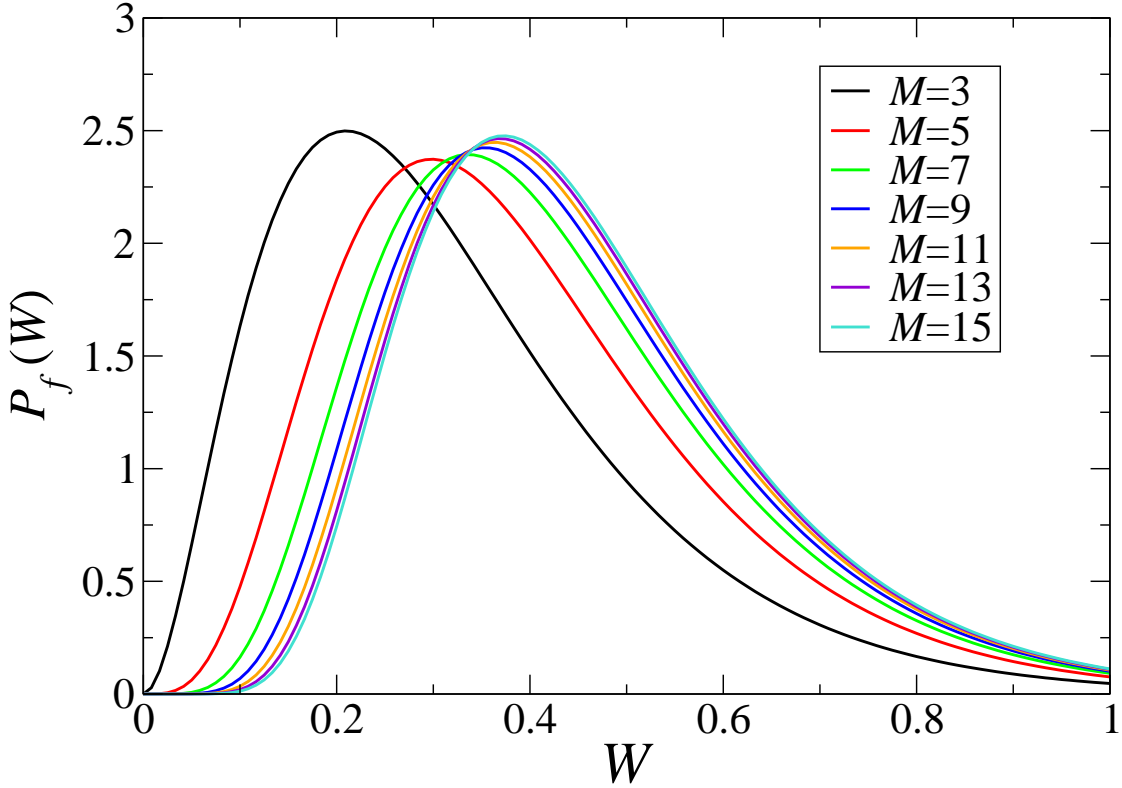}}
\caption{(Color online) Convergence of the quantum forward work distributions for a
         quasi-static switching between the potentials in
         Fig.~\ref{fig:switch-oscillator}, generated from Eq.~\eqref{PW}
         using the bead dispersion relation
         \eqref{dispersion-beads}. Plotted is $P_f(W)$ for various
         values of $M$.}
\label{fig:shape-first}
\end{figure}

The asymptotic convergence of the cumulants is different in the bead
regularization scheme, as is evident when the sum in Eq.~\eqref{cums} is first split
up into a sum from $k=-k^*$ to $k^*$ and a sum of modes with
$k^*\leq|k|\leq k_c=M/2$, and the latter is approximated by an
integral:
\begin{align}
   \kappa_j &
   \sim C_j +
    2(j-1)!
      \left(\frac{\Delta\omega^2}{2\beta}\right)^j
      \int_{k^*}^{M/2} \!
      \frac{\mathrm dk}{\omega_k^{2j}}.
\end{align}
where $C_j$ is the contribution of the $k<k^*$ modes. If
$k^*=\order{\hbar\beta\sqrt{\Delta\omega^2}}$, $C_j$ can be shown to
converge quickly as $M\to\infty$.  Using Eq.~\eqref{dispersion-beads}, one
gets
\begin{align}
   \frac{\kappa_j -C_j}
    {2(j-1)!}
   &\sim 
      \left(\frac{\beta\hbar^2\Delta\omega^2}{8M^2}\right)^jM
      \int_{\frac{k^*}{M}}^{1/2} \!
      \frac{\mathrm dq}{\sin^{2j}(\pi q)}.
\label{integral-form}
\end{align}
where the integration variable was changed to $q=k/M$.
The integral can be performed using
\begin{align*}
  \int^x \frac{\mathrm dx'}{\sin^{2j}x'} =  
-
\sum_{i=0}^{j-1}
\frac{(2j-2)!!(2i-1)!!}{(2j-1)!!(2i)!!}\frac{\cos x}{\sin^{2i+1} x}
,
\end{align*}
leading to
\begin{align}
   \frac{\kappa_j -C_j}
    {2(j-1)!}
   &\sim 
      -\left(\frac{\beta\hbar^2\Delta\omega^2}{8\pi^2}\right)^j
\frac{1}{2j-1}
\frac{1}{k^{*2j-1}}
+ \order{M^{-2}}.
\label{asymptote-beads}
\end{align}
Thus all cumulants in the bead regularization converge as $1/M^2$.
Note that this behaviour is consistent with the numerical results presented in
Fig. 3 of the preceding paper\cite{preceding}.

Even though all cumulants converge in the same way, the coefficients
in front of the $1/M^2$ terms often turn out to get smaller for higher
order cumulants, so that the shape will still appear to converge rather
quickly, as Fig.~\ref{fig:shape-first} illustrates: for large enough
$M$, the shapes of the forward work distributions for different values
of $M$ are very similar, but shifted along the $W$ axis.

Since the cumulants of the work distribution converge in the limit
$M\to\infty$ in both regularization schemes, the work distribution
itself is well defined and converges as $M^{-1}$ or $M^{-2}$ for the
Fourier and the bead regularizations, respectively, in spite of the
fact that the average energy $\langle H\rangle$ diverges linearly with
$M$.

\subsection{Convergence of the Jarzynski relation}

We will end this section with an explicit demonstration that
Jarzynski's non-equilibrium work relation is obeyed and converges to
the correct quantum result in the limit
$M\to\infty$. First note from Eq.~\eqref{generating0} that the left-hand side
of Eq.~\eqref{JE} may be reformulated as $\langle \exp(-\beta
W)\rangle_{\lambda_A}=G(i\beta)$. Using Eqs.~\eqref{Gu}, \eqref{Gnu} and
\eqref{andef}, one finds
\begin{align}
  \left\langle e^{-\beta W}\right\rangle
  &= \frac{\omega_A}{\omega_B} \prod_{k=1}^{k_c}
  \frac{\omega_k^2+\omega_A^2}{\omega_k^2+\omega_B^2}.
\label{product2}
\end{align}
This product coincides with the product on the right-hand side of
Eq.~\eqref{product}, showing that the Jarzynski equality $\left\langle
e^{-\beta W}\right\rangle=e^{-\beta\Delta F}$ holds and that the
resulting free energy difference converges to the exact result in the
limit $M \rightarrow \infty$ in the same way, i.e., as $1/M$ for the
Fourier regularization and as $1/M^2$ for the bead regularization, as
expected from Eqs.~\eqref{JEcum}, \eqref{asymptote-Fourier} and
\eqref{asymptote-beads}.

\section{Finite switching speed}
\label{finiteswitching}

\subsection{Work distribution}

Consider now the case in which the switch from $\omega_A$ to
$\omega_B$ is done in a finite (fictitious) time $\tau$, using the
protocol
\begin{equation}
   \omega^2(t) = \omega_A^2+\Delta\omega^2\frac{t}{\tau},
   \label{finite-switch}
\end{equation}
where $\Delta \omega^2$ was defined in Eq.~\eqref{delta-omega-def}, so that
$\omega^2(\tau)=\omega_B^2$.

As in the quasi-static process, the Fourier modes are independent and
can be handled separately using a time dependent frequency
$\Omega_k(t)$ defined through Eqs.~\eqref{Omegandef} and
\eqref{finite-switch}. The solution of Eqs.~\eqref{eom1} and \eqref{eom2} for
a set of initial conditions $\mathbf X_k(0) =(\tilde{x}_k(0),
\tilde{p}_k(0))$ for this switching protocol can be written as a
linear mapping
\begin{equation}
   \mathbf X_k(t) = \mathsf M_k(t) \mathbf X_k(0),
   \label{linearMapping}
\end{equation}
where $\Det\mathsf M_k(t) = 1$ since the dynamics conserves phase
space volume.  The mapping matrix $\mathsf M_k(t)$ can be written
explicitly in terms of the Airy functions\cite{AbramowitzStegun}
$\phi_1(t) = \Ai ( -\Omega_k^2(t)/b)$ and $\phi_2(t) = \Bi (
-\Omega_k^2(t)/b)$ with $b=\left( |\Delta \omega^2|/\tau \right)^{2/3}$
as
\begin{equation}
\mathsf M_k(t) = \tilde{\mathsf M}_k(t) \tilde{\mathsf M}^{-1}_k(0)
\end{equation}
where
\begin{equation}
   \tilde{\mathsf M}_k(t) = 
      \begin{pmatrix}
        \phi_1(t)         & \phi_2(t) 
        \\
        m\dot{\phi}_1 (t) & m \dot{\phi}_2 (t)
      \end{pmatrix} 
   .
\end{equation}
which yields
\begin{widetext}
\begin{align}
   \mathsf M_k(t) = 
      \pi \begin{pmatrix}
             \Ai(-y_t)\Bi'(-y_0)-\Bi(-y_t)\Ai'(-y_0) 
             &
             \frac{\sigma}{m\sqrt b}[\Ai(-y_t)\Bi(-y_0)-\Bi(-y_t)\Ai(-y_0)]
             \\
             \sigma m\sqrt b\,[\Bi'(-y_t)\Ai'(-y_0)-\Ai'(-y_t)\Bi'(-y_0)]
             &
             \Bi'(-y_t)\Ai(-y_0)-\Ai'(-y_t)\Bi(-y_0)
      \end{pmatrix}
   ,
\end{align}
\end{widetext}
where $\sigma=\pm 1$ depending on the sign of $\Delta\omega^2$,
$y_t=\Omega_k^2(t)/b$, and $\Ai'$ and $\Bi'$ are the derivatives of
the Airy functions.

Defining a diagonal matrix
\begin{equation*}
   \mathsf D_k(t) = 
      \begin{pmatrix}
         \frac12m \Omega_k^2(t) & 0 
         \\
         0                     & \frac{1}{2m}
      \end{pmatrix}
   ,
\end{equation*}
the energy contribution of mode $k$ at time $t$ can be written
as
\begin{align}
   H_k(t) 
      & =
      \mathbf X^T_k(t)\mathsf D_k(t)\mathbf X^T_k(t)
=
\mathbf X^T_k(0)\mathsf H_k(t)\mathbf X^T_k(0)
\end{align}
where the ``Hamiltonian'' matrix is given by
\begin{equation}
  \mathsf H_k(t) = \mathsf M^T(t)\mathsf D_k(t) \mathsf M(t).  
\label{Ham-matrix}
\end{equation}
{}From Eq.~\eqref{Hkdef}, the work contribution of mode $k$ can then be
written as
\begin{equation*}
W_k = \mathbf X_{k}^T(0)\left[ \mathsf H_k(\tau) - \mathsf
  D_k(0)\right] \mathbf X_k(0)
\end{equation*}
where we used that $\mathsf H_k(0)=\mathsf D_k(0)$.  It is now
straightforward to compute the generating function $G_k(u)$ of $W_k$,
\begin{align}
 G_k(u) &=
 \frac{\int \mathrm d\mathbf X_k(0) \:
          e^{ \mathbf X_k^T(0) \left[ \mathrm iu \mathsf
 H_k(\tau)-(\beta+\mathrm iu)
\mathsf D_k(0)\right] \mathbf X_k(0)}}
{\int \mathrm d\mathbf X_k(0) \:
e^{-\beta \mathbf X_k^T(0) \mathsf D_k(0) \mathbf X_k(0)}}
\nonumber \\
&= \left[\Det\Big(\mathsf I - \frac{\mathrm iu}{\beta}\mathsf
C_k\Big)\right]^{-1/2} ,
\label{generatingFunction}
\end{align}
where $\mathsf I$ is the identity matrix and
\begin{equation}
   \mathsf C_k 
      = \mathsf D^{-1/2}_k(0)\mathsf H_k(\tau)\mathsf D^{-1/2}_k(0)
        - \mathsf I
   .
   \label{Ckdef}
\end{equation}

To illustrate the dependence of the work distributions on $\tau$,
we need to perform the Fourier inverse on $G(u)=\prod_k G_k(u)$. For
$|k|>0$, this is simple, since $k$ and $-k$ are degenerate and yield
the combined contribution
\begin{equation}
G_k^2(u) =
\frac{1}{\Det\big(\mathsf I - \frac{\mathrm iu}{\beta}\mathsf
  C_k\big)}
  =\frac{1}{1-iu\frac{\lambda_k^{(1)}}{\beta}}
  \frac{1}{1-iu\frac{\lambda_k^{(2)}}{\beta}},
\label{Gk1}
\end{equation}
where $\lambda_k^{(1)}$ and $\lambda_k^{(2)}$ are the two eigenvalues
of $\mathsf C_k$. The right hand side of Eq.~\eqref{Gk1} has the form of two
independent exponential modes with rates
$\beta/\lambda_k^{(1,2)}$. The Fourier inverse of $\sum_{0<|k|\leq
k_c} G_k(u)$ is therefore
\begin{equation}
    \tilde P(W)=\Theta(\gamma'_0W) \sum_{k=1}^{2k_c}
           \frac{|\gamma'_k|}
          {\prod_{\stackrel{\mbox{\scriptsize$q=1$}}{q\neq k}}^{2k_c}
          \big(1-\frac{\gamma'_k}{\gamma'_q}\big)}
          e^{-\gamma'_kW},
\label{PtildeW}
\end{equation}
where by definition $\lambda_k^{(i)} = \beta/\gamma'_{2(k-1)+i}$. This
expression does not contain the contribution of the mode $k=0$, for
which
\begin{equation*}
G_0(u) = \frac{1}{\sqrt{\Det\big(\mathsf I - \frac{\mathrm iu}{\beta}\mathsf
  C_0\big)}}
  =\frac{1}{\sqrt{(1-iu/\gamma_+)(1-iu/\gamma_-)}},
  \end{equation*}
  where $\gamma_+=\beta/\lambda_0^{(1)}$ and
  $\gamma_-=\beta/\lambda_0^{(2)}$, and whose Fourier inverse is given by
  \begin{equation}
    P_0(W) = \sqrt{\gamma_+\gamma_-}
      e^{-\frac{\gamma_++\gamma_-}{2}W}I_0\left(\frac{\gamma_++\gamma_-}{2}W\right) ,
\label{P0W}
\end{equation}
where $I_0$ is the zeroth order modified Bessel function of the first
kind. The derivation of this result, which holds when $\gamma_+$ and $\gamma_-$ are both
positive, is given in Appendix \ref{appB}. In fact it can be shown
that for monotonically increasing $\omega$, such as in
Eq.~\eqref{finite-switch}, $\gamma_+$ and $\gamma_-$ must always be positive,
though this restriction need not hold for more complicated protocols. 

\begin{figure}[t]
\centerline{\includegraphics[width=0.9\columnwidth]{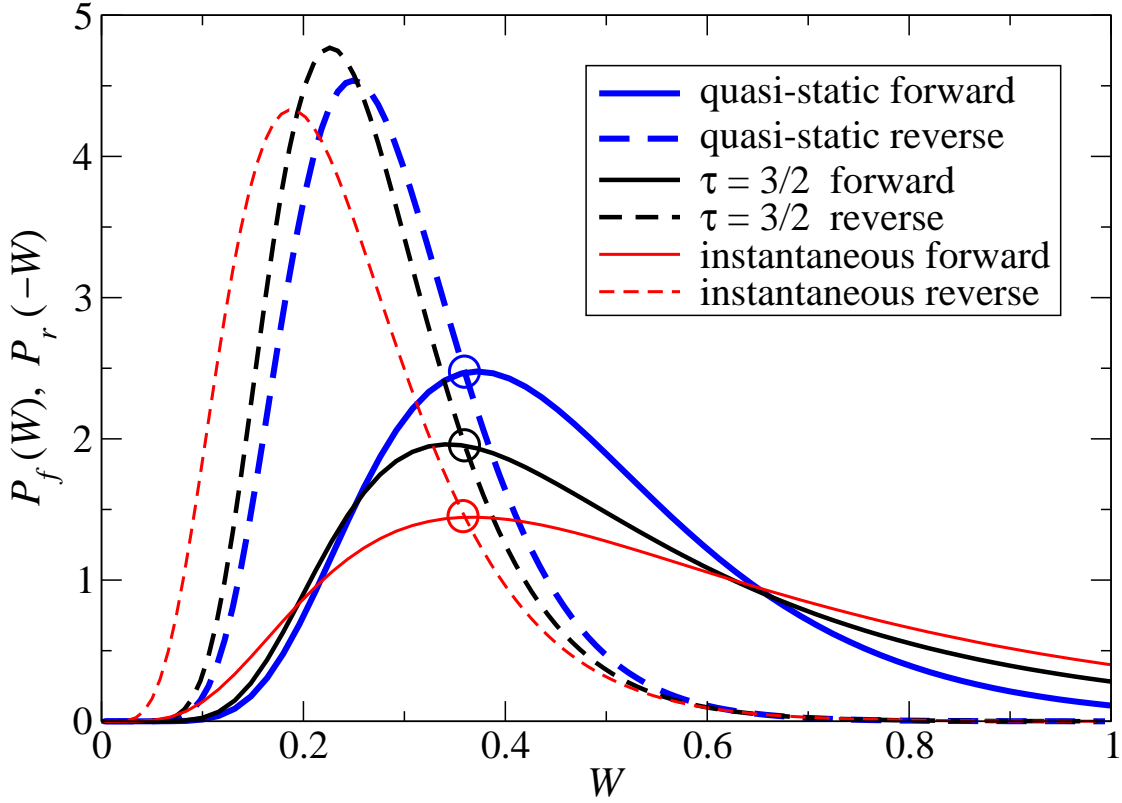}}
\caption{(Color online) Forward and reverse quantum work
         distributions for various switching speeds between the
         harmonic potentials in Fig.~\ref{fig:switch-oscillator},
         generated from Eq.~\eqref{PW} for the quasi-static case, from
         Eq.~\eqref{workPW} for the finite $\tau=3/2$, and from
         Eq.~\eqref{instantaneous} for the instantaneous case. Note
         that the intersection points occur at the same value of $W$,
         as they must according to the Crooks fluctuation relation.}
\label{tau-dependence}
\end{figure}

The complete work distribution function is therefore
\begin{equation}
        P(W) = \int_0^W \!\mathrm dW'\:P_0(W')\tilde P(W-W'),
\label{workPW}
\end{equation}
which unfortunately does not lead to a closed form, but can easily be
computed numerically since $P_0$ and $\tilde P$ are known
analytically.  As an example, the resulting $P(W)$ for the same
parameters as for the quasi-static case is shown in
Fig.~\ref{tau-dependence} for $\tau=3/2$, both for the forward and
reverse process.

\subsection{Convergence of the work distribution}

To determine the convergence properties of the work distribution as
$M\to\infty$, we will once again use the cumulants of the work.  From
the generating function in Eq.~\eqref{generatingFunction}, we have
\begin{align}
\sum_{j=1}^{\infty}
\frac{(\mathrm iu)^j}{j!}
\kappa^{(k)}_j &= 
- \frac{1}{2} \Tr\log\Big(\mathsf I - \frac{\mathrm iu}{\beta} \mathsf C_k\Big).
\end{align}
By expanding the logarithm and equating like powers of $u$ for 
$j \geq 1$, one finds the cumulants of the different modes to be 
\begin{equation}
  \kappa_j^{(k)} = \frac{(j-1)!}{2\beta^j} \Tr \mathsf C_k^j  .
  \label{kappa-j-finite}
\end{equation}
One thus has for the cumulants of the total work
\begin{equation*}
\kappa_j = \frac{(j-1)!}{2\beta^j} \sum_{|k|\leq k_c}\Tr \mathsf C_k^j.
\end{equation*}
This is of similar form to the expression \eqref{cums} for the work
cumulants in the quasi-static process, which could be shown to
converge because of the asymptotic property of $\gamma_k$ as
$k\to\infty$ in Eq.~\eqref{gamma-asymptotics}. A similar asymptotic property
here has to involve the two eigenvalues $\lambda_k^{(1)}$ and
$\lambda_k^{(2)}$ of the real symmetric $2\times2$ matrix $\mathsf
C_k$ for large $k$. This asymptotic analysis starts with the behavior
of the Airy functions for large negative
arguments\cite{AbramowitzStegun}
\begin{subequations}
\begin{align}
   \Ai(-y) 
      & \sim
        \frac{\sin\big(\frac23y^{3/2}+\frac{\pi}{4}\big)}{\pi^{1/2}y^{1/4}}
\\
   \Bi(-y) 
      & \sim
        \frac{\cos\big(\frac23y^{3/2}+\frac{\pi}{4}\big)}{\pi^{1/2}y^{1/4}}
\\
   \Ai'(-y) 
      & \sim
        -\frac{\cos\big(\frac23y^{3/2}+\frac{\pi}{4}\big)}{\pi^{1/2}}y^{1/4}
\\
   \Bi'(-y) 
      & \sim
        \frac{\sin\big(\frac23y^{3/2}+\frac{\pi}{4}\big)}{\pi^{1/2}}y^{1/4}
,
\end{align}
\end{subequations}
with corrections of relative $\order{y^{-3/2}}$. Using these
asymptotic expressions, one can write for the mapping matrix needed in
Eq.~\eqref{Ckdef}
\begin{equation}
   \mathsf M_k(\tau) 
    \sim 
        \begin{pmatrix}
            \frac1\eta_k\cos\theta_k
            &
            \frac{1}{\eta_k m\Omega_k(0)}\sin\theta_k
            \\
            -\eta_k m\Omega_k(0)\sin\theta_k
            &
            \eta_k\cos\theta_k
        \end{pmatrix}
   \label{Mk-asymptotic}
   ,
\end{equation}
with $\theta_k=\frac23(y_\tau^{3/2}-y_0^{3/2})$ and
$\eta_k\equiv\sqrt{\Omega_k(\tau)/\Omega_k(0)}$.  Substituting this
$\mathsf M_k(\tau)$ into Eq.~\eqref{Ham-matrix} for $\mathsf H_k$ using
$\Omega_k(\tau)=\eta_k^2\Omega_k(0)$ in $\mathsf D_k(\tau)$, and
substituting the result into expression \eqref{Ckdef} for $\mathsf
C_k$, yields
\begin{equation}
  \mathsf C_k \sim (\eta_k^2-1)\mathsf I.
\label{Ckresult}
\end{equation}
Thus it is clear that asymptotically, both eigenvalues of $\mathsf
C_k$ are equal to $\lambda^{(1)}_k=\lambda^{(2)}_k=\eta_k^2-1$. But
$\eta_k^2-1$ is precisely equal to $\beta/\gamma_k$, so that
Eq.~\eqref{kappa-j-finite} gives
\begin{equation*}
  \kappa_j^{(k)} = \frac{(j-1)!}{\beta^j} \Big(\frac\beta{\gamma_k}\Big)^j
                 = \frac{(j-1)!}{\gamma_k^{j}},
\end{equation*}
which precisely coincides with the asymptotic expression of the
cumulants of mode $k$ in the quasi-static case, Eq.~\eqref{cums}. In
retrospect, this is not too surprising: higher $k$ modes have higher
frequencies $\Omega_k$, so that the relative switching speed
$1/(\tau\Omega_k)$ decreases for increasing $k$, resulting in a
quasi-static behavior for the work distributions of the high-$k$
modes.

Since the behavior of the work cumulants for large $k$ is the same as
in the quasi-static switching, the cumulants themselves converge for
the finite switching speed as well, and in exactly the same way, i.e.,
$\order{1/M^{2j-1}}$ in the Fourier regularization and $\order{1/M^2}$
for the bead regularization.  However it should be stressed that the correspondence
between the quasi-static and finite-switching case only holds for the
large $k$ modes, and that the lower $k$ modes do differ between the
two cases and the total work distribution can vary substantially
depending on the switching speed.

\subsection{Convergence of the Jarzynski relation}

With these results for the generating function, the Jarzynski equality
\eqref{JE} can be once more checked. Setting $u=i\beta$ in
Eq.~\eqref{generatingFunction}, the generating function gives
\begin{align}
 G_k(i\beta)
 &=
 \left( \frac{\Det\mathsf D_k(0)}{\Det\mathsf H_k(\tau)}\right)^{1/2}
 = \left( \frac{\Det\mathsf D_k(0)}{\Det\mathsf D_k(\tau)}\right)^{1/2}
 \nonumber\\&
 = \frac{\Omega_k(0)}{\Omega_k(\tau)},
\label{Gkib}
\end{align}
where the fact that $\Det \mathsf M(\tau)=1$ has been used.  Putting
together all the contributions to $G(i\beta)$, one obtains the
same expression as for the quasi-static case, Eq.~\eqref{product2}.
The subsequent calculation therefore applies, showing that the Jarzynski
relation, Eq.~\eqref{JE}, in the form of Eq.~\eqref{Jarztest}, also 
holds for finite switching rates.

The result in Eq.~\eqref{Jarztest} holds not only for any $\tau$ using the
protocol in Eq.~\eqref{finite-switch}, but in fact for {\it any protocol}. A
different Hamiltonian dynamics would only change the linear mapping
$\mathsf M_k(\tau)$, which would still obey $\Det\mathsf M_k(\tau)=1$,
since this follows purely from the Hamiltonian nature of the
dynamics. Thus, Eq.~\eqref{Gkib} would still hold, from which the Jarzynski
relation follows.

\section{Instantaneous switching}
\label{instantaneousswitching}

While the work distribution for finite $\tau$ could only be expressed
analytically up to a convolution, for $\tau=0$, it is possible to
derive a fully analytical form. In this instantaneous limit, the
system does not have time to change its positions or momenta, whence
$W_k=\frac12m\Delta\omega^2|\tilde x_k|^2$. This allows the generating
function $G_k(u)$ to be computed, leading to
\begin{align}
  G(u) = 
\frac1{\sqrt{1-\mathrm i u\frac{\Delta\omega^2}{\beta\omega^2}}}
\prod_{k=1}^{k_c}\frac1{1-\mathrm i
  u\frac{\Delta\omega^2}{\beta\Omega_k^2}}.
\label{Gofuinstantaneous}
\end{align}
The generating function can be inverted, to yield
\begin{align}
  \tilde P(W)=\Theta(\gamma''_0W) \sum_{k=1}^{k_c} 
       \frac{|\gamma''_k|\erfi\big(\sqrt{(\gamma''_k-\gamma''_0)W}\big)}
        {\sqrt{\frac{\gamma''_k}{\gamma''_0}-1}
          \prod_{\stackrel{\mbox{\scriptsize$q=1$}}{q\neq k}} ^{k_c}
        \big(1-\frac{\gamma''_k}{\gamma''_q}\big)}
       e^{-\gamma''_kW},
\label{instantaneous}
\end{align}
where $\gamma''_k=\beta\Omega_k^2/\Delta\omega^2$ and
$\erfi(x)=-\mathrm i\erf(\mathrm ix)$ is the complex error function.
This work distribution $P(W)$ has also been plotted in Fig.~\ref{tau-dependence}, for
the same parameters as for the quasi-static and finite-$\tau$ cases
and both for the forward and reverse process. One sees that regardless
of the speed of the process, the crossing points $W_c$ of the
forward and reverse distributions are all the same, and equal to the free
energy difference.

To check the convergence of the Jarzynski equality in the
instantaneous switching case,
one substitutes $u=\mathrm i\beta$ into Eq.~\eqref{Gofuinstantaneous}. For each
of the modes, the generating function at $u=\mathrm i\beta$ coincides
with the result for the finite switching speed, i.e.\ the right-hand side of
Eq.~\eqref{Gkib}. As a result, the convergence of the free energy as computed
from the Jarzynski equality in the instantaneous case is the same as it
was for the finite switching case, as would be expected since the
former is the $\tau\to0$ limit of the latter.

The convergence of the work distribution might be expected to
follow similarly by taking the limit $\tau\to0$ of the cumulants found in the 
case of finite switching, but it turns out
that the limits $\tau\to0$ and $k\to\infty$ do not
commute. The physical reason is that no matter how fast the finite
switching is, for fixed positive definite $\tau$ and growing $k$, there are always $k$
modes whose frequencies $\Omega_k$ are faster than the switching rate
$\tau^{-1}$, and for those modes the switching is nearly quasi-static. But
if $\tau=0$, then the process is instantaneous for all modes, regardless of
their $k$ value. 

To see the noncommutativity of the limits mathematically, note
that for instantaneous process in which the positions and momenta
of the system do not have time to change, the $\mathsf M_k(\tau)$
matrix is equal to the identity matrix. The matrix $\mathsf C_k$
defined in Eq.~\eqref{Ckdef} then assumes the form
\begin{align}
  \mathsf C_k =
  \begin{pmatrix}\frac{\Omega_{kB}^2}{\Omega^2_{kA}}-1&0\\0&0\end{pmatrix},
\label{Ckresult0}
\end{align}
where $\Omega_{kA}^2=\omega_k^2+\omega_A^2$ and
$\Omega_{kB}^2=\lim_{\tau\to0}\Omega^2_k(\tau)=\omega_k^2+\omega_B^2$.
This form is not the same as the $\tau\to0$ limit of Eq.~\eqref{Ckresult}.
{}From Eqs.~\eqref{kappa-j-finite} and \eqref{Ckresult0}, one finds for the
mode cumulants in the instantaneous switching case
\begin{align}
  \kappa_j^{(k)} &=
  \frac{(j-1)!}{2\beta^j}\left(\frac{\Omega_{kB}^2}{\Omega_{kA}^2}-1\right)^{j}
\nonumber\\
 &= \frac{(j-1)!}{2\gamma^j_k}\left(\frac{\Omega_{kB}}{\Omega_{kA}}+1\right)^{j},
\end{align}
where Eq.~\eqref{andef} was used. For large $k$,
$\Omega_{kB}/\Omega_{kA}+1\to2$, so that $\kappa_j^{(k)}\propto
\gamma_k^{-j}$ in this regime.  Apart from a $k$ independent prefactor, the
asymptotic behavior of the cumulants as a function of $k$ is the same
as in the previous cases, so that the convergence of the work
distribution is also the same.

\section{Comparison with real time quantum dynamics}
\label{realquantum}

The viability of the non-equilibrium path-integral approach for the
computation of quantum free energy differences is made possible by the
use of a fictitious dynamic, thus avoiding the problems associated
with the description of the real time quantum evolution of the system,
in contrast to other extensions of the Jarzynski and Crooks
fluctuation relations to the realm of quantum mechanics, based on the
operator formulation of quantum
dynamics\cite{Mukamel1,Mukamel2,Maes,Talkneretal07b,Talkneretal07a,TalknerHanggi07,DeffnerLutz08}.
\begin{figure}[t]
\centerline{\includegraphics[width=0.9\columnwidth]{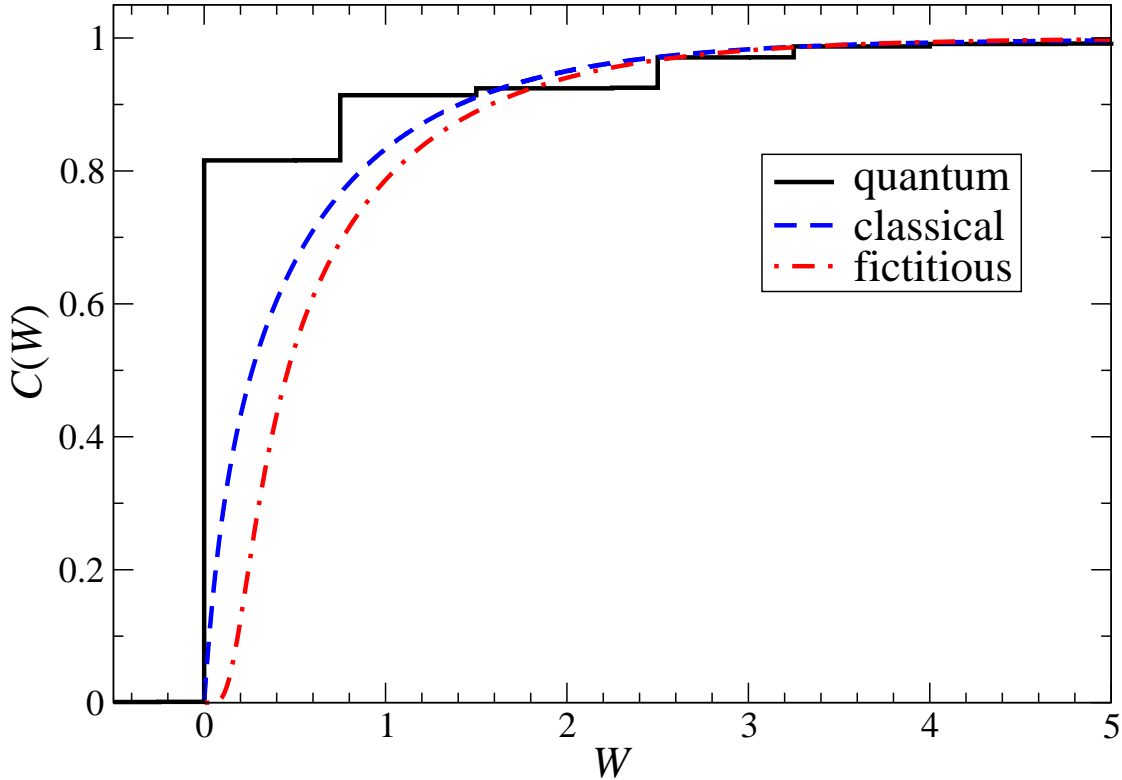}}
\caption{(Color online) Comparison of the work distributions in the non-equilibrium
  process of switching the oscillator strength linearly from
  $\omega_A=1/2$ to $\omega_B=5/4$ in a time $\tau=3/2$, starting in
  canonical equilibrium with inverse temperature $\beta=4$, for
  quantum, classical, and and fictitious dynamics.\label{comparison}}
\end{figure}
As a consequence, however, the work computed within this scheme bears
no relation to the work performed in any real quantum process
except in the classical limit $\hbar\beta\to0$, where there are
no contributions to the work distribution from the non-zero $k$ modes
in the fictitious dynamics. The work distribution for an isolated
quantum system in which the frequency is changed in real (rather than
fictitious) time has recently been worked out by Deffner and
Lutz\cite{DeffnerLutz08}, so that a direct comparison between the
fictitious dynamics and real quantum dynamics is possible.

The work distribution for the quantum harmonic oscillator consists of
a sum of delta functions, since the real quantum work distribution can
be obtained by summing over all possible transitions with the
appropriate transition amplitudes worked out by Husimi\cite{Husimi53}.
Thus, for the purpose of comparison, it is better to consider the
cumulative distribution function (CDF) $C(W)=\int_{-\infty}^W\!\mathrm
dw\:P(w)$\cite{DeffnerPrivate}.  Figure~\ref{comparison} shows the
comparison between the classical CDF [cf.~Eq.~\eqref{P0W}], the
quantum CDF (using Husimi's transition amplitudes) and the CDF
resulting from the fictitious path dynamics
[cf.~Eqs.~\eqref{PtildeW}--\eqref{workPW}] for $\omega_A=1/2$,
$\omega_B=5/4$, $\tau=3/2$ and $\beta=4$.  The three curves are
obviously very different. The quantum distribution is composed of
steps, while the other two are continuous. The quantum distribution
also does not agree with the fictitious dynamics on
average. Interestingly, the deviation of the real quantum dynamics
from the classical case is less then the deviation from the fictitious
dynamics. Another difference between the real quantum dynamics and the
dynamics in the other cases is the non-zero probability of negative
work values predicted by the distribution, which, as suggested in
Ref.~\onlinecite{DeffnerLutz08}, is due to excited states decaying to
lower states. This negative tail disappears in the classical limit for
the monotonic protocol in Eq.~\eqref{finite-switch}. The negative
tails in the classical limit found in Ref.~\onlinecite{DeffnerLutz08}
can only exist for non-monotonic protocols.

Given these observations, it is clear that the fictitious dynamics is
very different from real quantum dynamics and has little direct
physical content.

\section{Conclusions}
\label{conclusions}

The non-equilibrium methods for the calculation of free energy
differences in quantum systems in the context of the path integral
representation of the canonical partition function presented in the
previous paper were applied to the harmonic oscillator to show that
the work distribution function is well-defined as the regularization
parameter $M$ is taken to infinity. Instead of using the real quantum
dynamics of the system, the path integral representation allows a
fictitious path to be defined for which the Jarzynski and Crooks
relations are valid. By evolving the ring polymer in the path integral
representation under fictitious dynamics, the difficulties associated
with the complexity of the full evolution of a quantum system are
avoided.

In particular, expressions for the distribution $P(W)$ of the work $W$
done in the non-equilibrium fictitious process in which the strength
changes linearly during a time $\tau$, were derived, in the form of a
single convolution for finite $\tau$, and in fully explicit form for
$\tau\to0$ (the instantaneous limit) and $\tau\to\infty$ (the
quasi-static limit). From $P(W)$, it was shown that the Jarzynski
relation holds for this case for any dynamic switching process based
on (isolated) Hamiltonian dynamics.  The convergence of the resulting
free energy difference was obtained for both regularizations, and goes
as $\order{M^{-1}}$ for the Fourier regularization and as
$\order{M^{-2}}$ for the bead regularization. The nature of the
convergence of the cumulants of $P(W)$ as $M\to\infty$ was also
determined for any $\tau$. Whereas the beads regularization leads to
cumulants which all converge as $M^{-2}$, the $j$th cumulant converges
as $\order{M^{1-2j}}$ in the Fourier representation, implying that the
shape of the distribution converges faster than its position along the
$W$ axis.  However, one can use perturbative arguments with the
harmonic oscillator as the zeroth order system to show that the work
distributions in the Fourier and bead regularization converge as $1/M$
and $1/M^2$, respectively. Indeed, in the path integral simulations
using the bead regularization presented in the preceding paper\cite{preceding},
one sees a $1/M^2$ convergence for the free energy difference as well
as for the first and second cumulant of the work distribution
function. Given the analytical proof given in this paper and the
numerical evidence in the preceding paper, it can be
expected that the convergence of the method is general.

Other regularization schemes based on the splitting method were
also briefly considered and it was found that splitting schemes 
optimized for molecular dynamics need not  be optimal for the
convergence of path integrals, in contrast to higher order splitting
schemes which will have better asymptotic convergence properties.

The difference between the fictitious dynamics and real quantum
dynamics was demonstrated by direct comparison of the work
distribution. Not only is the nature of the distribution different
(delta peaks for the quantum case, a smooth function for the
fictitious dynamics), but the two also do not agree on average.
Nonetheless, the free energy found from the non-equilibrium method
with fictitious dynamics is the exact quantum free energy~difference.

\acknowledgments

We would like to thank Sebastian Deffner for helpful correspondence.
The authors would like to acknowledge support by grants from the
Natural Sciences and Engineering Research Council of Canada NSERC.

\appendix

\section{Energy conservation of the classical harmonic oscillator under the
  HOA2 integration scheme}
\label{appA}

The HOA2 scheme is a second order integrator for molecular dynamics
simulations, like the Verlet scheme, but contains a parameter $\eta$
which can be tuned to make the dynamics as accurate as
possible\cite{Omelyanetal02a,Omelyanetal02,Omelyanetal03}. In this
appendix, the level of energy conservation as a function of $\eta$
will be determined explicitly for the classical harmonic oscillator.

The state of the classical harmonic oscillator is represented by a
position $x$ and momentum $p$. Combining these quantities into a
two-dimensional vector $\Gamma=(\omega x,p/m)$, the energy of the
oscillator is given by $E=\frac{m}{2}|\Gamma|^2$.

In applying the HOA2 splitting scheme in Eq.~\eqref{hoa2} to molecular
dynamics, the kinetic operator $\hat T$ is to be replaced by the free
Liouville operator $\mathcal L_T=\{p^2/2m,.\}$ and the potential
operator $\hat V$ by the interaction Liouville operator $\mathcal
L_V=\{V(r),.\}$, where $\{ \cdot, \cdot \}$ is the Poisson bracket
operator. The sum of these two Liouville operators is the full
Liouvillian $\mathcal L=\mathcal L_T+\mathcal L_V$. Because of the
linearity of the equations of motion of the classical harmonics
oscillator, the exponentials of these two Liouville operators can be
written as the matrices
\begin{align}
e^{\mathcal L_T t} = \mathsf A(t)&=\begin{pmatrix}1&\tau\\0&1\end{pmatrix}\\
e^{\mathcal L_V t} = \mathsf B(t)&=\begin{pmatrix}1&0\\-\tau&1\end{pmatrix},
\end{align}
respectively, which act on the vector $\Gamma$, and where $\tau=\omega
t$. For the one-step propagator, one thus finds from Eq.~\eqref{hoa2}
\begin{align}
e^{\mathcal L t/M} 
&\approx
\mathsf B\Big(\frac{\eta t}{M}\Big)
\mathsf A\Big(\frac{t}{2M}\Big)
\mathsf B\Big(\frac{(1-2\eta)t}{M}\Big)
\mathsf A\Big(\frac{t}{2M}\Big)
\mathsf B\Big(\frac{\eta t}{M}\Big)
\nonumber\\&=
\begin{pmatrix}
1-\frac{\tau^2}{2M^2}+\frac{\eta(1-2\eta)\tau^4}{4M^4}
&
\frac{\tau}{M}-\frac{(1-2\eta)\tau^3}{4M^3}
\\
-\frac{\tau}{M}+\frac{\eta(1-\eta)\tau^3}{M^3}-\frac{\eta^2(1-2\eta)\tau^5}{4M^5}
&
1-\frac{\tau^2}{2M^2}+\frac{\eta(1-2\eta)\tau^4}{4M^4}
\end{pmatrix} .
\end{align}
We will denote this approximate propagator-matrix by $\mathsf P(t/M)$.
The approximate propagator $\mathsf P(t)$ over a time $t$ is given by
the $M$th power of $\mathsf P(t/M)$, which can be evaluated by
diagonalization:
\begin{align}
  \mathsf P\Big(\frac{t}{M}\Big) = \mathsf
  U\cdot\mathrm{diag}(\mu_1,\mu_2)\cdot\mathsf U^{-1},
\end{align}
with $\mu_1$ and $\mu_2$ the eigenvalues of $\mathsf P(t/M)$, so that
\begin{align}
  \mathsf P(t) = \mathsf
  U\cdot\mathrm{diag}(\mu_1^M,\mu_2^M)\cdot\mathsf U^{-1}.
\label{Peigen}
\end{align}
Because we want to minimize the error in front of the leading
correction term, we only need the expansions of the eigenvalues and of
$\mathsf U$ in inverse powers of $M$. The eigenvalues may be expressed
as
\begin{align*}
  \mu_1 &= 
1+\frac{\mathrm{i}\tau}{M}-\frac{\tau^2}{2M^2}-
\frac{\mathrm{i}(1+2\eta-4\eta^2)\tau^3}{8M^3}+\mathcal O\Big(\frac{1}{M^{4}}\Big)
\\
\mu_2&=
1-\frac{\mathrm{i}\tau}{M}-\frac{\tau^2}{2M^2}+\frac{\mathrm{i}(1+2\eta-4\eta^2)\tau^3}{8M^3}+\mathcal
O\Big(\frac{1}{M^{4}}\Big) ,
\end{align*}
whence
\begin{align*}
  \mu_1^M &= 
e^{\mathrm i\tau}\left[1+\frac{\mathrm i(1-6\eta+12\eta^2)\tau^3}{M^2} +\mathcal O\Big(\frac{1}{M^3}\Big)\right]
\\
\mu_2^M&=
e^{-\mathrm i\tau}\left[1-\frac{\mathrm
i(1-6\eta+12\eta^2)\tau^3}{M^2} +\mathcal
O\Big(\frac{1}{M^3}\Big)\right] ,
\end{align*}
while the matrix $\mathsf U$ can be written as
\begin{align*}
\mathsf U &=
\begin{pmatrix} 
-\mathrm i+\frac{\mathrm i(1-6\eta+4\eta^2)\tau^2}{8M^2} 
&
\mathrm i-\frac{\mathrm i(1-6\eta+4\eta^2)\tau^2}{8M^2} 
\\1&1
\end{pmatrix}
+\mathcal O\Big(\frac{1}{M^3}\Big).
\end{align*}
Substituting these expressions into Eq.~\eqref{Peigen}, the deviations
in the total energy from its initial value as a function of the
initial conditions $x_0$ and $p_0$ can be written as
\begin{align}
  E(t) -E(0)&= \frac12m|\mathsf P(t)\cdot\Gamma_0|^2-\frac12m|\Gamma_0|^2
\nonumber\\&=
\frac{(1-6\eta+4\eta^2)\tau^2\sin^2\tau}{8M^2}
(m\omega^2x_0^2-p_0^2/m-2\omega x_0 p_0\cot\tau)
+\mathcal O(M^{-4}).
\label{conservation}
\end{align}
Thus, the leading violation of energy conservation is of order $1/M^2$
for general $\eta$. In particular, the Verlet scheme, for which
$\eta=1/4$, has second order violations in the energy
conservation. Equation \eqref{conservation} shows that the leading
order violation can be eliminated altogether by taking the solution of
$1-6\eta+4\eta^2=0$, i.e., $\eta=(3\pm\sqrt5)/4$. The larger of these
solutions leads to negative time steps in the HOA2 scheme, which can
lead to instabilities, so the value of $\eta$ to take is
$\eta=(3-\sqrt 5)/4=0.1909830056250526\ldots$.  This value of $\eta$
makes the scheme pseudo-fourth order for the harmonic oscillator, and
is very close to the general optimized value of
$\eta=0.1931833275037836\ldots$\cite{Omelyanetal02a}, which has been
demonstrated to have smaller variations in the total energy than the
Verlet scheme for general potentials in numerical
simulations\cite{VanZonetal08a}.

\section{Fourier inverse for the classical work distribution}
\label{appB}

\begin{figure}[t]
\centerline{\includegraphics[width=\columnwidth]{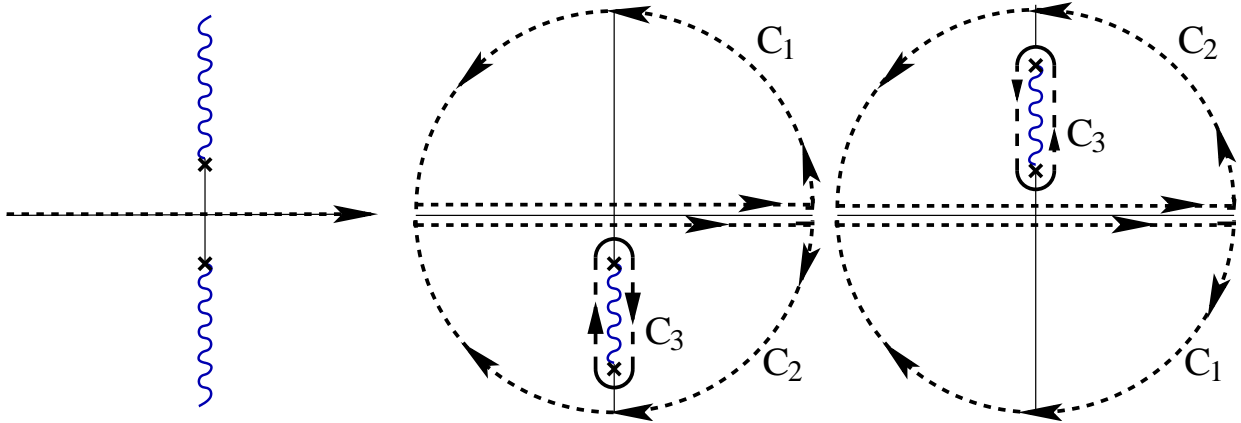}}
\caption{Integration contours (dotted lines) and branch cuts of
  $G_0(u)$ (wiggly lines) for different signs of $\gamma_+$ and
  $\gamma_-$, for the three cases that need to be distinguished in
  the evaluation of the Fourier inverse in the appendix. The crosses
  are the singular branch points.}
\label{contour}
\end{figure}

To arrive at the classical result Eq.~\eqref{P0W} for the finite-$\tau$
switching process, one needs to compute
\begin{align}
  P_0(W)=\frac{1}{2\pi}\int_{-\infty}^\infty\!\mathrm du \frac{e^{-\mathrm{i} uW}}
          {\sqrt{(1-\mathrm{i} u/\gamma_+)(1-\mathrm{i} u/\gamma_-)}}.
\label{P0WW}
\end{align}
The integrand is a two-valued complex function with two singularities
$-\mathrm{i} \gamma_{\pm}$, which are also the branch points. The
branch to take in the integral should be that for which the integrand
is $1$ at $u=0$ [since $G_0(0)=\int\mathrm dWP_0(W)=1$]. In addition,
no branch cut should be crossed as one integrates from $-\infty$ to
$+\infty$, i.e., the branch cut should not cross the
real axis. This restricts the choice of where to put the branch cut:
if $\gamma_+$ and $\gamma_-$ are of opposite sign, with $\gamma_+>0$
say, then in order to avoid a branch cut on the real axis the branch
cut should have two parts, one extending from $-i\gamma_+$ to
$-\mathrm i\infty$ and one from $\mathrm i|\gamma_-|$ to $i\infty$, as
in the left panel of Fig.~\ref{contour}. If, on the other hand,
$\gamma_+$ and $\gamma_-$ are both positive, then the branch points
$-\mathrm i\gamma_{\pm}$ both lie on the negative imaginary axis, and
it is convenient to put the branch cut between the two branch points,
as indicated in the middle panel of Fig.~\ref{contour}, while if
$\gamma_+$ and $\gamma_-$ are both negative, the situation is mirrored
with respect to the real axis, as shown in the right panel of
Fig.~\ref{contour}. This third case can also be treated by setting
$W\to-W$ and $u\to-u$ in Eq.~\eqref{P0WW}, and therefore does not need to be
treated separately.

For the first case, i.e., $\gamma_+$ and $\gamma_-$ of opposite
sign, with $\gamma_-$ chosen negative, one can shift the integration
line up or down without hitting the branch points, until the
integration line passes straight through the middle of the two branch
points.  In the integral in Eq.~\eqref{P0WW}, this corresponds to a
shift of the integration variable over $\mathrm
i(\gamma_++\gamma_-)/2$. Performing in addition a scaling,
such that $u=[-\mathrm{i}(\gamma_++\gamma_-)+t|\gamma_+-\gamma_-|]/2$,
gives
\begin{align}
  P_0(W)&=
   \frac{\sqrt{\gamma_+|\gamma_-|}}{2\pi}
        e^{-\frac{\gamma_++\gamma_-}2W}
\int_{-\infty}^\infty\!\mathrm dt \frac{e^{-\mathrm{i} \frac{|\gamma_+-\gamma_-|}{2}Wt}}
          {\sqrt{1+t^2}}
\nonumber\\
&=
   \frac{\sqrt{\gamma_+|\gamma_-|}}{\pi}
        e^{-\frac{\gamma_++\gamma_-}2W}K_0\Big(\frac{|(\gamma_+-\gamma_-)W|}{2}\Big),
\end{align}
using formula 8.432.5 in Gradshteyn and Ryzhik with $\nu=0$ and
$z=1$\cite{GradshteynRyzhik}.  This result was also found by Deffner
and Lutz [Eq.~(27) in Ref.~\onlinecite{DeffnerLutz08}] when they
considered the classical limit of a real quantum dynamical
process. However, $\gamma_+$ and $\gamma_-$ can only have opposite
signs if the oscillator frequency is changed non-monotonically, which
is not the case considered in the text.

If $\gamma_+$ and $\gamma_-$ are both positive, which occurs for
monotonically increasing oscillator frequencies
[cf.~Eq.\eqref{finite-switch}], one cannot simply shift the
integration line to run in between the branch points since one of the
branch points would be crossed. The integral is then evaluated as
follows. For $W<0$, one can construct a contour C$_1$ composed of
the real axis and an infinite semi-circle in the upper half of the
complex plane, as shown in Fig.~\ref{contour}. Because for $W<0$, the
factor $e^{-\mathrm{i} u W}$ decays exponentially when $u$ has a large
positive imaginary component, the semi-circle does not contribute to the integral, and
the contour integral over C$_1$ has the same value as the original
integral. Since no singularities lie within this contour, the integral
is zero, proving that for $W<0$, the integral is zero. If, on the
other hand, $W$ is positive, then the integrand decays for $u$ with a
large negative imaginary component, and the integral can be replaced
by an integration over the closed contour C$_2$ found by adding an
infinite semi-circle in the lower half of the complex plane, cf.\
Fig~\ref{contour}. Note that because of the location of the branch
cut, one remains on the same branch of the function going along C$_2$,
which is a requirement for the contour to be truly closed. The contour
C$_2$ does contain singularities, and in particular, it contains the
branch cut. The contour can be deformed without crossing singularities
to the contour C$_3$ in Fig.~\ref{contour} which goes around the
branch cut. One easily shows that the contributions from the parts
that go around the branch points vanish. Furthermore, the integrand
changes sign across the branch cut and the contour is traversed in
opposite directions on either sides, so that the contributions from
the left and the right segments of $C_3$ are equal, and the integral
becomes
\begin{align}
  P_0(W)=\int_{-\mathrm{i}\gamma_+}^{-\mathrm{i}\gamma_-} \!\mathrm du\frac{e^{-\mathrm{i} uW}}
          {\pi\sqrt{(1-\mathrm{i} u/\gamma_+)(1-\mathrm{i} u/\gamma_-)}}.
\end{align}
Changing integration variables to $t$ where
$u=[-\mathrm{i}(\gamma_++\gamma_-)+\mathrm{i}t(\gamma_+-\gamma_-)]/2$  reduces
this integral to 
\begin{align}
  P_0(W)=\frac{\sqrt{\gamma_+\gamma_-}}{\pi}
        e^{-\frac{\gamma_++\gamma_-}2W}
        \int_{-1}^1 \!\mathrm dt
          \frac{e^{-\frac{\gamma_+-\gamma_-}{2}Wt}}{\sqrt{1-t^2}}
\end{align}
Given the representation of the modified Bessel function of the first
kind given by formula 8.431.1 in Gradshteyn and Ryzhik with
$\nu=0$\cite{GradshteynRyzhik}, one finds Eq.~\eqref{P0W}.

\end{document}